\documentclass[12pt,a4paper]{article}
\usepackage{amssymb}%
\usepackage[T2A]{fontenc}
\usepackage[cp866]{inputenc}
\usepackage[russian,english]{babel}
\usepackage{graphicx}
\usepackage[dvips]{psfrag}
\normalbaselines
\usepackage[unicode]{hyperref}
\begin{document}
\begin{center}
{\Large How to detect the lightest glueball}\\[3mm]
{B.~P.~Kosyakov${}^{a,b}$, E.~Yu.~Popov${}^a$, and  M. A. Vronski{\u\i}${}^{a,c}$}\\[3mm]
{{\small ${}^a$Russian Federal Nuclear Center--VNIIEF, 
Sarov, 607188 Nizhni{\u\i} Novgorod Region;\\
${}^b$Moscow Institute of Physics {\&} Technology, Dolgoprudni{\u\i}, 141700 Moscow Region;\\ 
${}^c$Sarov Institute of Physics {\&} Technology, Sarov, 607190 Nizhni{\u\i} Novgorod Region}\\
} 
\end{center}
\begin{abstract}
\noindent
{We suggest a 
procedure for detecting the lightest glueball in a head-on collision of photons whose 
center-of-mass energy is in the range  $1.3-2$ GeV.   
We use recent evidence for scattering of light by light in Large Hadron Collider experiments 
as a phenomenological basis for our suggestion. 
With this evidence, the cross section of the lightest glueball creation in $\gamma\gamma$
collisions is estimated to be  $\sim 60$ nb.
The predominant mode of the lightest glueball decay, predicted from the gauge/gravity duality, 
proves to be the decay into two neutral vector mesons $\rho^0\rho^0$.
Because the $\rho^0$ decays into $\pi^+\pi^-$, a drastic increase of the $\pi^+\pi^-\pi^+\pi^-$ 
yield is expected as the center-of-mass energy approaches the value of the lightest glueball mass.
This fact  will be the unique signature of the lightest glueball detection.}
\end{abstract}


\newpage
\section{Introduction}
\label
{Introduction}
The prediction of hadrons containing no quarks, composed of gluons alone, was made 
\cite{FritzschGell-Mann}--\cite{JaffeJohnson} at the dawn of the age of quantum chromodynamics (QCD), 
the modern theory of the strong interactions.
In the past half-century, prodigious experimental efforts went into searching such particles, now 
known as glueballs.
However,  glueballs have not yet been observed with certainty \cite{Crede, Olive}.
Why so?
Are glueballs absent from Nature altogether?
Another belief prevails currently: the glueball field mixes with the 
quark-antiquark fields, $\left(u{\bar u}+d{\bar d}\right)/\sqrt{2}$ and $s{\bar s}$, to form 
the experimentally observed meson resonances \cite{Anisovich95}--\cite{Ochs}.

Nevertheless, there are a few dissenters from the view that the mixing of glueballs and quarks is 
unavoidable. 
For example, it was stated in \cite{KabanaMinkowski}--\cite{Vento} that the depth of quark-gluon 
plasmas, formed in heavy-ion collisions, may have a beneficial effect on the creation of pure 
ground state glueballs.
To verify this idea, we should be capable of distinguishing the possible yield of glueballs among 
many thousands of foreign tracks related to the explosion products of a quark-gluon plasma lump.
This is a challenge.    
The multiplicities of glueballs in a single central heavy-ion collision is estimated 
by fitting the hadron ratios observed in ${\rm Pb}-{\rm Pb}$ collisions at various 
energies in the Large Hadron Collider (LHC) to be $1.5-4$ glueballs \cite{Mishustin}.

In the present paper, we propose an alternative procedure for detecting pure ground state glueballs. 
Our concern here is with the lightest glueball, a color singlet of two gluons, specified by zero 
total angular momentum and positive parity and charge parity, $J^{PC}=0^{++}$. 
We denote this particle by ${\mathbb G}$. 
It was anticipated in the pioneer studies \cite{FritzschGell-Mann}--\cite{JaffeJohnson} that a 
scalar glueball can decay into two photons with opposite polarizations.  
The reverse process provides us with a way for creating pure ground state glueballs: 
$\gamma\gamma\to{\mathbb G}$. 
We explicate this idea, endeavor to justify it phenomenologically and show its feasibility in 
Sect.~\ref{Suggestion}.
The central theoretical problem though is to predict the decay products of ${\mathbb G}$.
Section \ref{Critique} is devoted to this problem.

Unlike quarks, gluons are uncharged, and have vanishing weak hypercharge and isospin 
couplings, so that ${\mathbb G}$ is immune from the electromagnetic and weak interactions. 
We thus expect that the strong interactions are responsible for the decay of ${\mathbb G}$.
This is tantamount to stating that the lightest glueball is subject to a deconfinement through its 
splitting into two gluons.
Now we come into a low-energy domain in which the QCD running coupling constant $\alpha_s(\mu)$ is 
large, and the perturbation technique is inoperative.
Three most-used ways for the description of phenomena in this domain are: 

(1) semiclassical approach, 

(2) lattice QCD simulations,  

(3) gauge/gravity duality.

In the settled semiclassical picture, quarks are represented by point particles linked together
by thin tubes which enclose the total color flux of the gluon field.
Nucleons are colorless objects composed of three valent quarks.
They are assumed to be assembled in nuclei by a residual color interaction similar to the multipole
van der Waals force between neutral molecules.
Meanwhile the capability to give a precise meaning  to the notion of the residual color interaction
turns out to be rather problematic \cite{KPV15}. 
That is why the Yukawa mechanism, thought of as a meson exchange mechanism, refined by several 
technical innovations, such as spontaneously broken chiral symmetry, effective
Lagrangians, and derivative expansions, still forms the basis for modern nuclear physics 
\cite{Weinberg1990}--\cite{Machleidt}.  
And yet the issue of understanding nuclei in terms of quarks is high on the agenda of the QCD 
developments.
We do not dwell on the efforts to address this issue in the framework of different semiclassical 
approaches (bags, potential models, etc.), and refer the interested reader to \cite{KPV17}
and references therein.

The Coleman theorem \cite{Coleman} dramatically hampers semiclassical treatments of glueballs.
By this theorem, no localized (kink-like) finite-energy solutions of pure Yang--Mills theory is 
available.
Strictly speaking, there is no rendition of a glueball as a localized object; we have not the 
foggiest notion of what the size and structure are peculiar to the lightest glueball.
For lack of color fields that issue out of fixed points and cancel each other, the notion of 
residual color interactions becomes quite farfetched.
We thus have to state that the lightest glueball does not interact with its environment until 
it splits into two gluons.
  
According to lattice and sum rule calculations, the lightest glueball has mass in the range of 
about $1.3-2$ GeV \cite{Anisovich95}--\cite{Ochs}. 
Furthermore, lattice QCD predicts the glueball mass spectrum.
However, the glueball coupling with ordinary hadrons hitherto eluded reliable analyzing.

In this paper, the decay of ${\mathbb G}$ is examined in the context of gauge/gravity duality, aka 
the correspondence between a quantum theory of gravity in anti-de Sitter space and conformal field 
theories in Minkowski space, AdS/CFT, and the holographic principle \cite{Maldacena}--\cite{Gubser}; for 
a full coverage of ideas and methods of gauge/gravity duality see \cite{Ammon, Nastase}. 
Loosely speaking, gauge/gravity duality is a doctrine whereby a good part of subnuclear physics in a 
four-dimensional realm is modelled on physics of black holes and similar objects (black rings, 
black branes, etc.) in five-dimensional anti-de Sitter space, ${\rm AdS}_5$, whose boundary is just 
this four-dimensional realm.
However, this understanding of gauge/gravity duality apparently indulges in wishful thinking.
In their popular science paper \cite{KlebanovMaldacena} Klebanov and Maldacena remind the reader of 
the physics joke about the spherical cow as an idealization of a real one, and admit that ``in the 
AdS/CFT correspondence, theorists have really found a hyperbolic cow''.
To remedy the situation, a major portion of the standard holographic mapping is to be amputated.
 
The main stream in the high-energy-physics research today develops the idea that a black hole in ${\rm AdS}_5$ is
mapped holographically onto a quark-gluon plasma lump in a four-dimensional realm \cite{Arefeva}.
This realization of holography in the framework of the Bekenstein--Hawking thermal treatment of 
gravitational phenomena in ${\rm AdS}_5$ was offered in \cite{Herzog}.
The line of reasoning is as follows.
Let us compare the values that the gravitational action
\begin{equation}
I=-\frac{1}{2\kappa^2}\int d^5 x\,\sqrt{g}\left(R+\frac{12}{L^2}\right)
\label
{gravitation_action}
\end{equation}
takes when two solutions are substituted in it, one describing thermal ${\rm AdS}_5$ and the other 
 describing thermal ${\rm AdS}_5$ with a Schwarzschild black hole
\footnote{To eliminate divergences of the action (\ref{gravitation_action}) arising from these 
substitutions, the integral must be regularized by means of a cut-off, or using a nontrivial 
dilaton field expectation value \cite{Herzog}.}.
It then transpires that the former is less than the latter at $T<T_c$, where $T_c$ is the 
Hawking-Page phase transition temperature, and conversely, the latter is less than the former at 
$T>T_c$.
It follows that the thermal ${\rm AdS}_5$ becomes unstable at $T>T_c$ to yield the black hole 
formation.  
The holographic image of this process on the four-dimensional boundary of ${\rm AdS}_5$ is a QCD 
phase transition implying the quark-gluon plasma formation at a critical temperature associated 
with $T_c$.      

In an alternative approach \cite{SakaiSugimoto-I}--\cite{BruennerParganlijaRebhan}, black Dp-branes 
are mapped holographically onto a subnuclear realm in the confinement phase.
The set of significant entities suitable for the holographic mapping was defined in \cite{KPV19} to 
be that containing only extremal black holes and other extremal black objects, which correspond to 
their holographic counterparts, stable microscopic systems.      

These two seemingly incompatible concepts of gauge/gravity duality are in fact quite consistent.
The Bekenstein--Hawking thermal analysis of gravitational phenomena forms the basis of the former, 
whereas thermal treatments are unrelated to the latter because extremal black objects live in a 
cold realm, $T=0$; they do not experience Hawking evaporation.
Their associated holographic counterparts from different QCD phases may belong to distinct sectors 
of the holographic mapping.

In Sect.~\ref{Critique}, following the general ideology of \cite{KPV19}, which narrows the 
holographic context down to extremal black holes and their stable subnuclear counterparts, we yet
seek for extracting useful information from the frontier zone separating the domain in which
the holographic principle holds and that in which this principle fails.

\section{How to create the lightest glueball}
\label
{Suggestion}
\begin{figure}[htb]
\begin{center}
\phantom{m}\hskip-1cm{\includegraphics[width=0.42\textwidth,angle=180]{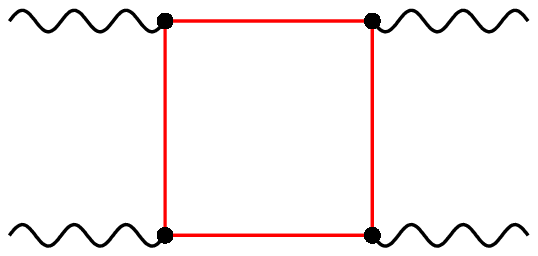}}
\hskip-2cm{\includegraphics[width=0.42\textwidth,angle=180]{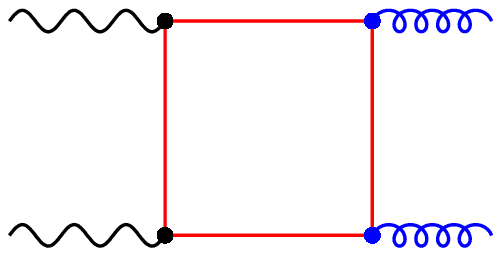}}
\hskip-2cm{\includegraphics[width=0.42\textwidth,angle=180]{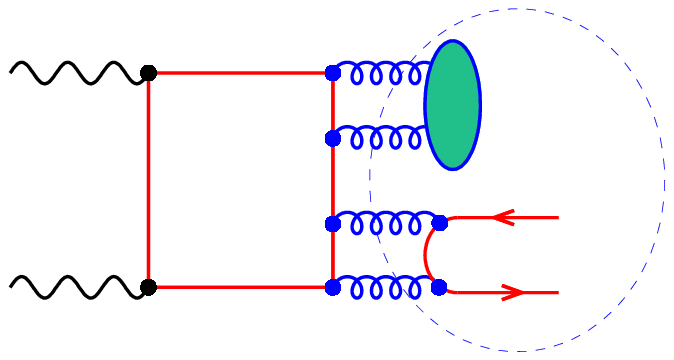}}
\begin{tabular}{p{0.3\textwidth}p{0.3\textwidth}p{0.3\textwidth}}
\parbox{0.3\textwidth}{\centerline{(a)}}&
\parbox{0.3\textwidth}{\centerline{(b)}}&
\parbox{0.3\textwidth}{\centerline{(c)}}
\end{tabular}
\caption{$\gamma\gamma$ collisions resulting in: (a) a two-photon system; (b) 
a two-gluon system; (c) an aggregate of a glueball and a meson}
\label{gamma-gamma}
\end{center}
\end{figure}

It was already noted that kinematics and symmetries allow the decay of ${\mathbb G}$ into two 
gamma-quanta with opposite spiralities.
The lower orders in $\alpha$ and $\alpha_s$ of this process is depicted in Figure~\ref{decay} as 
diagram $(c)$.   
The inverted diagram corresponds to the reverse of this process: a head-on $\gamma\gamma$ collision
at a center-of-mass energy $\sqrt s$ in the range $1.3-2$ GeV, with the helicity of the 
$\gamma\gamma$ system being zero, may result in the lightest glueball creation 
\footnote{We speak about this event in tentative modality because our concern is not with a
conversion $\gamma\gamma\to gg$, Figure~\ref{gamma-gamma}~(b), which can be accounted for in the
framework of the conventional perturbation theory, but with the occurrence of a bound state of two 
gluons.
Now we are not in a position to look into the formation of gluon confinement in detail, but some 
insight into this process will be gained from gauge/gravity duality in the next section.}. 

To test this idea, it is attractive to use a photon collider, the most extensively studied 
prospective device
\footnote{For a technical design report of the Photon Collider at TESLA see \cite{B}.} 
having its origin in the conversion of 
laser photons into high-energy gamma-quanta through the Compton scattering on 
high-energy electrons \cite{Ginzburg}.
The device consists of two beams of electrons moving towards each 
other to the interaction point ${\bf x}_\ast$.
The electrons collide with laser photons at a distance of about $1-5$ mm from ${\bf x}_\ast$.
After the scattering, the photons become gamma-quanta with energy comparable 
to that of the electrons and
follow their direction to ${\bf x}_\ast$ where they collide with similar 
counterpropagating gamma-quanta. 
The maximum energy $\omega$ of the gamma-quanta is given by
\begin{equation}
\omega= \frac{x}{x+1}\,E\,,
\quad
x\approx \frac{4E\omega_0}{m^2}\,, 
\label
{omega-max}
\end{equation}
where $E$ and $\omega_0$ are, respectively, the energy of the electrons and laser photons, and $m$ 
the electron mass. 
For example, $E=7.5$ GeV is needed if we are to convert the photon energy $\omega_0=1.17$
eV (Nd: glass laser) into the gamma-quantum energy $\omega=0.85$ GeV.

Using a laser with a flash energy of several joules one can obtain gamma-quanta whose spot size at 
${\bf x}_\ast$ will be almost equal to that of the electrons at ${\bf x}_\ast$, and the total 
luminocity of $\gamma\gamma$ collision will be comparable to the ``geometric'' luminocity of the 
electron beams.
The energy spectrum of the gamma-quanta becomes most peaked if the initial electrons are 
longitudinally polarized and the laser photons are circularly polarized. 
This gives almost a factor of 4 increase of the luminosity in the high-energy peak. 
The present laser technology has all elements needed for the required photon colliders \cite{B}.

In order to evaluate the feasibility of the conversion $\gamma\gamma\to gg$ in the discussed
layout, we invoke recent evidence for the scattering of light by light in quasi-real photon 
interactions of ultra-peripheral 
${\rm Pb}+{\rm Pb}$ collisions, with impact parameters larger than twice the 
radius of the nuclei, at a nucleon-nucleon center-of-mass energy $\sqrt{s}=5.02$ TeV by the 
ATLAS experiment at the LHC \cite{ATLAS_Collaboration, Aad}.
The fiducial cross section of the process 
${\rm Pb}+{\rm Pb}(\gamma\gamma)\to{\rm Pb}^{(\ast)}+{\rm Pb}^{(\ast)}\gamma\gamma$, for 
diphoton invariant mass greater than 6 GeV, is measured to be 
$78\pm 13 ({\rm stat}.)\pm 7 ({\rm syst}.)$ nb.
This result is in agreement with the Standard Model \cite{Enterria}--\cite{Bern}, where the
light-by-light scattering arises, in the leading order of $\alpha$, via one-loop diagrams, 
Figure~\ref{gamma-gamma} $(a)$.
An important point is that if we recalculate (according to \cite{Enterria}) the ATLAS experiment
result to the elementary cross section of the light by light scattering in vacuum at 
$\sqrt{s}=1.5$ GeV, we obtain $\sigma_{\gamma\gamma\to\gamma\gamma}\sim 70$ pb. 
 
Another piece of information derives from the usual QCD calculations of 
quarkonium partial widths \cite{ApelquistPolitzer, BarbieriGattoKogerler}.
The rule for changing two external photon lines by two external gluon lines
[see  Figure~\ref{gamma-gamma}, respectively, plots $(a)$ and $(b)$] 
refers to the factor
\begin{equation}
\frac{9}{8}\frac{\alpha_s^2(m_c)}{\alpha^2}\approx 845\,, 
\label
{to_gluon-to_photon}
\end{equation}
where $\alpha_s(\mu)$ stands for the QCD running coupling constant, which being taken at
$\mu$ equal to the mass of charmed quarks $m_c= 1.28$ GeV is  $\alpha_s\approx 0.2$.
Indeed, this value of  $\alpha_s$ can be obtained \cite{Apelquist, Vainshtein} from the ratio
\begin{equation}
\frac{\Gamma\left(J/\phi\to {\rm hadrons}\right)}{\Gamma\left(J/\phi\to e^+e^-\right)}=
\frac{5\left(\pi^2-9\right)\alpha_s^3}{18\pi\alpha^2}\,, 
\label
{Gamma_to_Gamma}
\end{equation}
whose experimental value is $\approx 10$.

All things considered, the cross section of the lightest glueball creation in 
head-on $\gamma\gamma$ collisions at energy region about $\sqrt s=1.7$  GeV 
is expected to be  $845\,\sigma_{\gamma\gamma\to\gamma\gamma}\approx 60$ nb. 

Note that it is the pure ground state glueball which will be produced 
in the proposed experiment; the creation of ${\mathbb G}$ is not obscured by the mixing effects 
with isoscalar $q{\bar q}$ mesons.  
To see this, we compare the probabilities of production of an unmixed scalar 
glueball and an aggregate of a glueball and a meson [Figure~\ref{gamma-gamma}, 
respectively, plots $(b)$ and $(c)$]: the latter, being represented by higher perturbative orders, 
is suppressed by a factor of  $\alpha^4_s\approx 1.6\cdot 10^{-3}$ as against the former . 
 
Finally, the fact that the Stanford Linear Collider luminocity, $\approx 3\cdot 10^{30} {\rm cm}^2/{\rm s}$, 
is four orders of magnitude greater than the luminocity $\approx 5\cdot 10^{26} {\rm cm}^2/{\rm s}$ 
of the light by light scattering event in ${\rm Pb} - {\rm Pb}$ collisions at the LHC 
\cite{ATLAS_Collaboration, Aad} counts in favor of feasibility of photon colliders with fairly
high luminocity. 
This is an added reason for the proposed experiment layout.  

\section{Decay of ${\mathbb G}$}
\label
{Critique}
We assume that the lightest glueball is unstable, and hence we would like to know its decay channels.
Our prime interest here is with the decay products of the predominant mode.
The detection of just this yield in the photon collider experiment may evidence 
that it is the lightest glueball which has been created in a $\gamma\gamma$ collision.  

The following criterion for discriminating between stable and unstable systems of the subnuclear 
world was offered in \cite{KPV19}: a system is stable when its gravitational dual is an 
extremal black object.
It may appear that the converse is also true, namely if a microscopic system is unstable, then its 
gravitational counterpart is an ordinary black hole amenable to Hawking evaporation.
But such is not the case.
An obstacle for gaining a well-defined correspondence between unstable microscopic systems and 
evaporating black holes relates to basic tenets of quantum mechanics, the principle of 
reversibility and the principle of identity and indistinguishability.

Quantum mechanical processes are reversible.
Suffice it to say that the probability amplitude for a decay equals that for the pertinent 
recombination.
By contrast, the black hole evaporation is an irreversible process.

All microscopic systems of the same species are identical and indiscernible.
Their gravitational duals must exhibit identical properties.
Suppose that a Lorentz frame ${\cal O}'$ moves past another Lorentz frame ${\cal O}$ at constant 
velocity $V$, and the standard synchrony of their clocks is established.
Let ${\cal O}$ and ${\cal O}'$ meet, and, at the instant of their meeting, both reset their clocks 
to $0$.
Suppose that ${\cal O}$ carries a black hole whose properties (mass, electric charge, and spin), 
at $t=0$, are identical to those of a further black hole assigned to ${\cal O}'$.
The rate of evaporation, as measured on the proper time, is common for both black holes.
Therefore, at $t>0$, the black holes possessed by ${\cal O}$ and ${\cal O}'$ have different masses; 
the relativistic effect of time dilation keeps evaporating black holes from being regarded as 
identical objects
\footnote{The reader will easily find this situation to be a kind of the well-known twin paradox 
in which the twins are the evaporating black holes, and the responsibility for their identification 
rests with the quantum mechanical principle of identity and indistinguishability of particles.}.

Our assumption that ${\mathbb G}$ is an unstable particle seems to forbid us from invoking  the 
holographic principle because this principle holds for stable systems of the microscopic world and fails for
unstable systems.
Nevertheless, there is a frontier zone in which the holography could be handled, with extreme care,
as a guiding principle, to clarify general features of the decay mechanism typical for unstable 
particles of this zone.
What are the reasons for this anticipation?
     
The decay of a particle may be attributed not only to an incessant evaporation process of a black 
hole but also to a single act of spontaneous splitting of the black hole into two or several black holes.  
If this splitting gives rise to extremal black holes
\footnote{This act may be thought of as a possible scenario for the completion of the history of an
evaporating black hole. 
For the scenarios proposed in the literature see \cite{Chen}.}, then the process is 
apparently reversible: isolated extremal black holes remain unchanged for any length of time, and
their collision can recover the initial black hole.  

We next imagine that a single black hole is in five-dimensional anti-de Sitter universe, or else, a 
black hole is widely separated from other matter, and can be considered to be almost free from 
force interactions. 
Then the issue of quantum mechanical identity of this black hole with objects of the same kind is of 
little significance.
A holographic counterpart of this black hole is a microscopic system which is exceptional in the 
sense that the vast majority or even all but one of ways for its decay are forbidden. 

We will focus upon the decay of neutral spinless particles whose gravitational duals are
taken to be Schwarzschild black holes.

Such black holes in ${\rm AdS}_5$ have the greatest possible spatial isometry group SO$(4)$, 
equivalent to SO$(3)\times$SO$(3)$, which is holographically mapped upon the SU$(2)_L\times$ SU$(2)_R$ 
chiral invariance of QCD with $N_f=2$ flavors.  
But chiral invariance is spontaneously broken in the confinement phase down to the isospin SU$(2)_V$ 
symmetry. 
Therefore, it is reasonable to expect that the dual isometry group SO$(4)$ is also broken down to SO$(3)$.
In other words, a {Schwarzschild black hole} in ${\rm AdS}_5$ is amenable to {spontaneous splitting} 
into black objects whose symmetry is limited to SO$(3)$, 
such as spinning black holes of Myers and Perry \cite{MP}
\footnote{An apparent objection to this statement is that the Schwarzschild geometry is stable 
against small perturbations in the classical context. 
However, the case in point is a quantum tunnelling of one black hole through the event horizon of
the other  black hole, a phenomenon which falls into the Hawking radiation pattern.}.

By the duality argument, a neutral spinless particle dual to this Schwarzschild black hole  
must decay into spinning particles.
We emphasize again that this rule is only valid for particles from the frontier zone.
The lightest neutral spinless meson 
$\pi^0$, whose weak and strong interactions are suppressed, decays through 
the electromagnetic channel as $\pi^0\to\gamma\gamma$. 
The lightest scalar particle immune from the electromagnetic and strong 
interactions, the Higgs boson $H^0$, decays into pairs of heavy fermions 
($b{\bar b}$, $\tau{\bar\tau}$) or gauge vector bosons ($W^+W^-$, $ZZ$, $gg$, 
$\gamma\gamma$).
The lightest glueball is likely to fall into the same category because the only conceivable way
for its decay is related to a splitting into two vector particles ${\mathbb G}\to gg$.

\begin{figure}[htb]
\phantom{m}\hskip-1cm{\includegraphics[height=4.5cm,width=0.4\textwidth,angle=180]{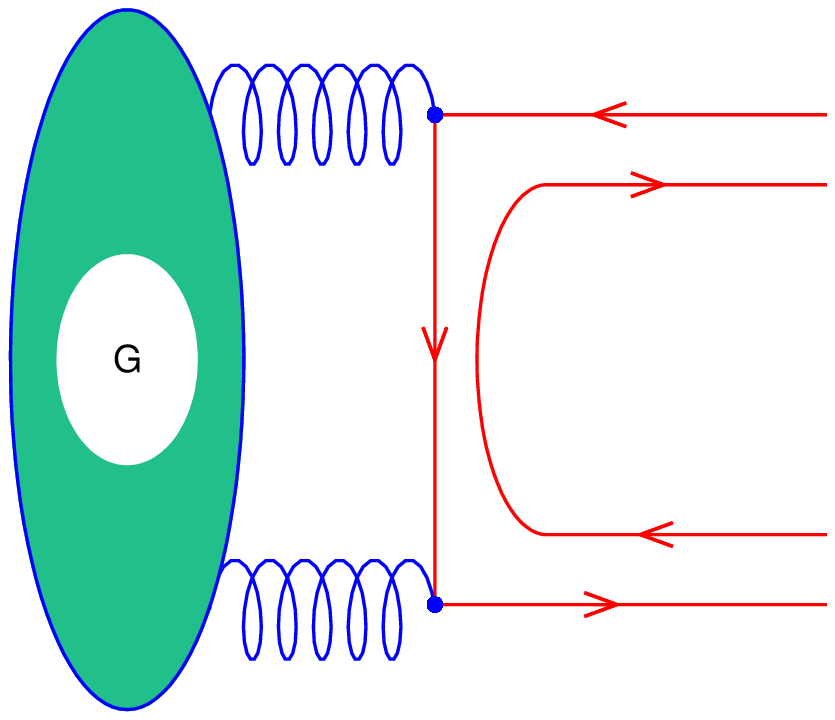}}
\hskip-2cm{\includegraphics[height=4.5cm,width=0.4\textwidth,angle=180]{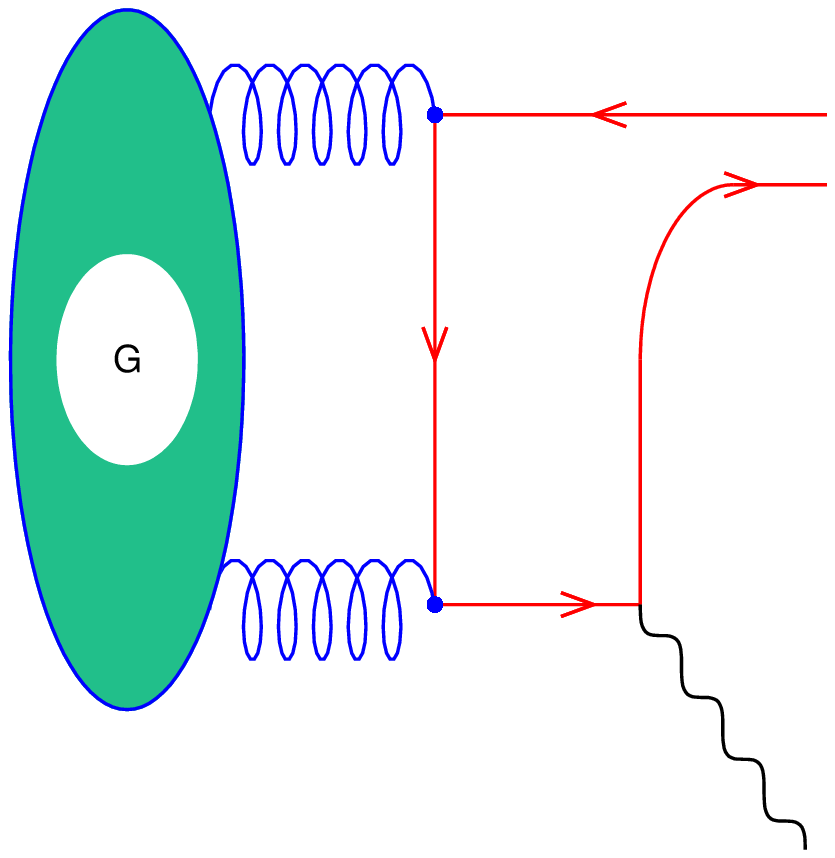}}
\hskip-2cm{\includegraphics[height=4.5cm,width=0.4\textwidth,angle=180]{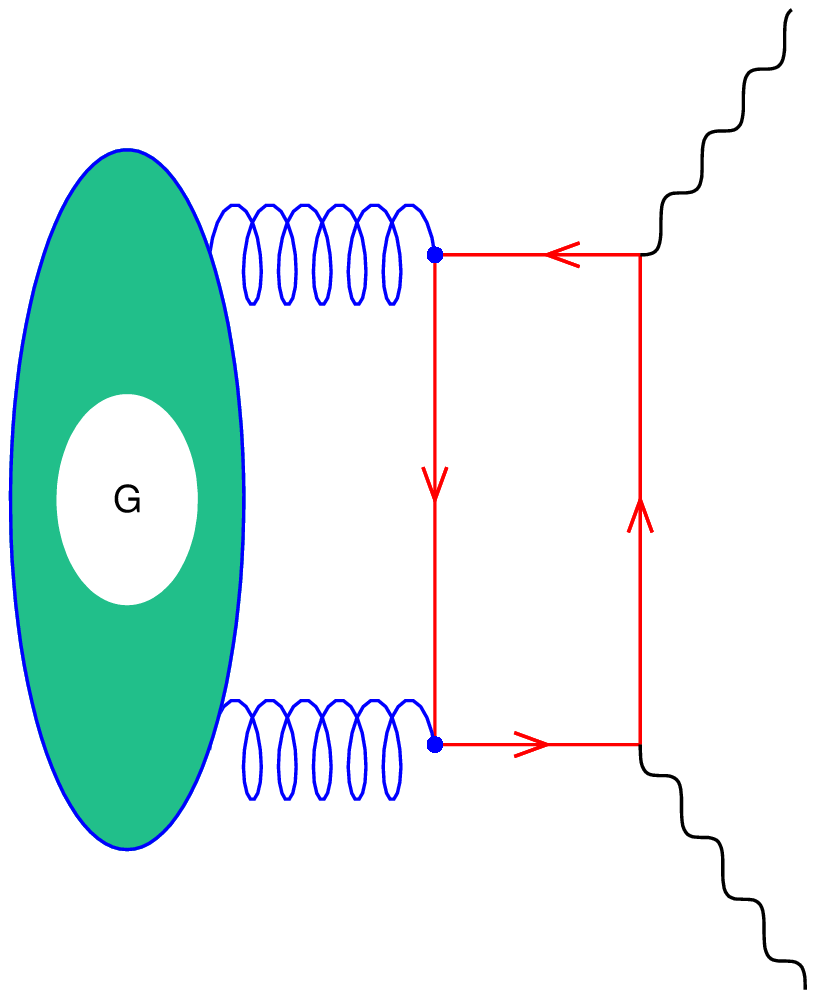}}
\begin{tabular}{p{0.3\textwidth}p{0.3\textwidth}p{0.3\textwidth}}
\parbox{0.3\textwidth}{\centerline{(a)}}&
\parbox{0.3\textwidth}{\centerline{(b)}}&
\parbox{0.3\textwidth}{\centerline{(c)}}
\end{tabular}
\caption{Decay modes for the lightest glueball}
\label{decay}
\end{figure}

Observable manifestations of this rule for the decay of ${\mathbb G}$ can be established in the
framework of the usual perturbation theory whose lower orders in  $\alpha$ and $\alpha_s$ are depicted 
in Figure~\ref{decay}.
Photons, quarks, and gluons are shown as the sine waves, oriented lines, and spirals, respectively. 
The outgoing vector mesons, composed of quarks and antiquarks, are represented 
by couples of antiparallel rays.
Diagram $(a)$ illustrates the decay mode whose outcome is a pair of truly neutral light-quark vector 
mesons, $\rho^0\rho^0$, or, alternatively, $\omega\omega$.
The masses of these particles are, respectively, $m_{\rho^0}=775$ MeV, and $m_\omega=783$ MeV.
Diagram $(b)$ displays the outcome as a photon and a truly neutral vector meson, which may be 
given by either $\rho^0$, or $\omega$, or $\phi$ ($m_\phi=1019$ MeV).
Diagram $(c)$ sketches a two-photon decay mode.
The ratio of probabilities of these modes can be roughly estimated as $1:{O}(\alpha):{O}(\alpha^2)$.
\psfrag{G}[c][c]{${\mathbb G}$}

It follows that the holographically inspired predominant decay mode of the lightest glueball
is ${\mathbb G}\to\rho^0\rho^0$.
With the fact that $\rho^0$ decays into $\pi^+\pi^-$ ($\approx 100${\%} fraction; $\Gamma=
149.1\pm 0.8$ MeV) \cite{Olive}, one may expect a drastic increase in the 
$\pi^+\pi^-\pi^+\pi^-$ yield as the center-of-mass energy $\sqrt s$ approaches the value of the 
lightest glueball mass $m_{\mathbb G}$, with the orbital momentum quantum number 
$l$ being equal to 1 for each $\pi^+\pi^-$ pair.

\section{Discussion}
\label
{Discussion}
\begin{figure}[htb]
\phantom{m}
\hskip-1cm{\includegraphics[width=0.4\textwidth,angle=180]{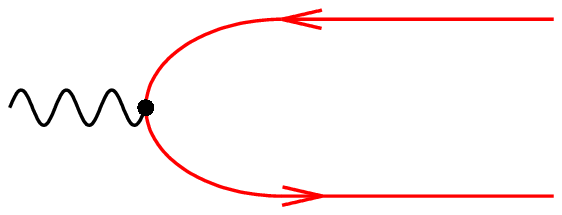}}
\hskip-2cm{\includegraphics[width=0.4\textwidth,angle=180]{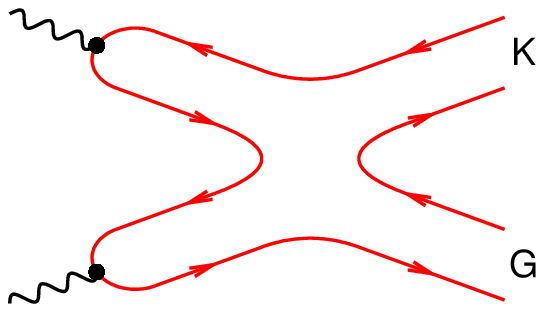}}
\hskip-2cm{\includegraphics[width=0.4\textwidth,angle=180]{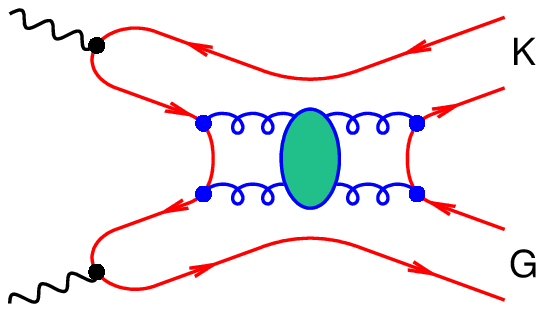}}
\begin{tabular}{p{0.3\textwidth}p{0.3\textwidth}p{0.3\textwidth}}
\parbox{0.3\textwidth}{\centerline{(a)}}&
\parbox{0.3\textwidth}{\centerline{(b)}}&
\parbox{0.3\textwidth}{\centerline{(c)}}
\end{tabular}
\caption{The strong interaction of photons: (a) a virtual photon becomes a 
vector meson; (b) two virtual photons collide to give a $K{\bar K}$ 
system; (c) two virtual photons collide to give a $K{\bar K}$ system 
through the mediation of a glueball.}
\label{virtual}
\end{figure}

The use of  $\gamma\gamma$ collisions for the possible yield of glueballs has already been realized 
as ingredients of experimental studies on $ee$ and $e{\bar e}$ collisions \cite{Acciarri}--\cite{Uehara}.
However, the physics behind those experiments differs from the physics behind the hunting of the 
lightest glueball proposed here. 
Unlike the former which has to do with virtual photons, the latter bears on real photons.
According to the concept of the vector meson dominance, virtual photons are capable of creating 
neutral vector mesons, Figure~\ref{virtual}~(a).
The most plausible scenario for the collision of such photons relates to the conversion of the 
created vector mesons into a pair of scalar mesons 
\footnote{Say into $K{\bar K}$, which is typical for the experiments analyzed in  
\cite{Acciarri}--\cite{Uehara}.}, represented by planar tree diagrams, Figure~\ref{virtual}~(b).
If higher order diagrams, with gluon lines being taken into account, the resonances appearing in 
the cross section may be attributed to the interposition of a glueball.
In fact, this effect is due to a glueball mixed with $q{\bar q}$ states. 
This comes into particular prominence from Figure~\ref{virtual}~(c): the glueball history is 
sandwiched between the quark and antiquark world lines.

On the other hand, the probability amplitude that a real photon spontaneously turns to a massive 
particle is very small, while a head-on collision of two such photons brings into existence of a 
massive entity ${\mathbb G}$.
If the total helicity of the colliding photons is zero, and the center-of-mass energy $\sqrt{s}$ 
equals the mass of the lightest glueball, this ${\mathbb G}$ proves to be just the lightest glueball. 
Although the outcome of the array of reactions completed in the above experiments, 
$\gamma\gamma\to\phi\phi\to K{\bar K}\to 4\pi$, is similar to that of the array 
of the proposed reactions, $\gamma\gamma\to {\mathbb G}\to\rho^0\rho^0\to 4\pi$, the 
contents of the compared processes seem much different.
\psfrag{K}[c][c]{$K$}
\psfrag{G}[c][c]{$\bar K$}

It might be well to mention a recent event \cite{Csorgo}, indirectly touching the subject of
our discussion, -- the experimental discovery of the odderon (a virtual colorless three-gluon state), 
which is likely to be of great importance in the dynamics of three-gluon vector glueball.
 
\section{Concluding remarks}
\label
{Conclusion}
Glueballs are enigmatic objects.
A major portion of their properties are associated with infrared effects which are beyond the 
control of the conventional perturbation theory.
The derivative expansion in chiral perturbations is also irrelevant here because the lightest 
glueball ${\mathbb G}$ is heavier than 1 GeV.

Semiclassical treatments are hampered by the fact that no rendition of a glueball is 
available due to the Coleman theorem \cite{Coleman}.  
We thus have not the slightest notion what the size and structure are peculiar to the lightest 
glueball.
Lattice QCD predicts the glueball mass spectrum.
However, the glueball coupling with ordinary hadrons still does not have a good lattice-based method 
of solution.

It remains to try to address the gauge/gravity duality or, what is the same,  the holography.
As was made clear in Sect.~\ref{Critique}, the gauge/gravity duality bears no relation to unstable 
microscopic systems.
However, there is a frontier zone separating the domain in which the holography holds and that in 
which it fails.
We argued that the holography can be used as a guiding principle to clarify qualitative properties 
of the decay mechanism typical for unstable particles from the frontier zone.
The lightest glueball is just among the set of particles from the frontier zone. 
The holographic reasoning together with calculations of $m_{\mathbb G}$ predict the predominant 
decay mode of  ${\mathbb G}$ to be ${\mathbb G}\to\rho^0\rho^0$.

In Sect.~\ref{Suggestion}, we proposed a procedure of detecting the lightest glueball in a head-on 
$\gamma\gamma$ collision, and adduced some phenomenological justifications.   
Our proposal holds much practical promise because the photon collider experiments are an extensively 
elaborated project, ready to be implemented (for the state of the art see, {\it e.~g}., \cite{B}).
The cross section of the process $\gamma\gamma\to{\mathbb G}$ estimated as $\sim 60$ nb is in
general consistent with the results of experiments for exploring the process 
$\gamma\gamma\to\rho^0\rho^0$ in the same energy range which had already been made in the 80s (see, 
{\it e. g.}, \cite{Kolanoski} and references therein).

In addition, the idea to detect ${\mathbb G}$ in $\gamma\gamma$ collisions has truly theoretical 
virtues.
There is good reason to believe that an unmixed scalar glueball will be created in the proposed 
experiment, which is confirmed by the estimated cross section of the ${\mathbb G}$ creation and 
feasible luminocity of photon colliders.
By now,  five isoscalar resonances are established: $f_0(500)$, $f_0(980)$, $f_0(1370)$, $f_0(1500)$, 
and $f_0(1710)$
\footnote{Note that these resonances decay into spinless (rather than spinning) particles $\pi\pi$, $K{\bar K}$, 
$\eta\eta$, $\eta\eta'$ \cite{Olive}.} whose masses are in the range of about $0.5-2$ GeV \cite{Olive}.
Among them, only $f_0(1710)$ can be suspected to be an unmixed scalar glueball \cite{Janowski, 
Albaladejo}, but this view was challenged \cite{Geng}.  
If the proposed experiment will exhibit a comparatively narrow state at the energy $\sqrt{s}$ other 
than the mass of any one of these resonances, this will strongly suggest that a new kind of 
hadron matter without quarks, the lightest glueball, has been discovered.   
A precise measurement of $m_{\mathbb G}$ will provide further impetus and guide the way for 
improving lattice QCD.  
The expected drastic increase in the $\pi^+\pi^-\pi^+\pi^-$ yield at $\sqrt{s}=m_{\mathbb G}$ 
will be a direct experimental evidence in support of the gauge/gravity duality prediction.

\section*{Acknowledgment}
\label
{Acknowledgment}
We are grateful to Masud Chaichian, Wolfgang Ochs, Valery Tel'nov, and Vicente Vento for useful 
discussions.

\end{document}